# TRIDENT SEGMENTATION CNN: A SPATIOTEMPORAL TRANSFORMATION CNN FOR PUNCTATE WHITE MATTER LESIONS SEGMENTATION IN PRETERM NEONATES

*Yalong Liu[1#], Jie Li[1], Miaomiao Wang[2#], Zhicheng Jiao[3], Jian Yang[2], and Xianjun Li[2*]*

[1] Visual Information Processing Lab, School of Electronic Engineering,
Xidian University, Xi'an 710071, China
[2] Department of Radiology, The First Affiliated Hospital of Xi'an Jiaotong University,
Xi'an 710061, China
[3] Perelman School of Medicine at the University of Pennsylvania, Philadelphia, PA 19104, USA

**ABSTRACT**

Accurate segmentation of punctate white matter lesions (PWML) in preterm neonates by an automatic algorithm can better assist doctors in diagnosis. However, the existing algorithms have many limitations, such as low detection accuracy and large resource consumption. In this paper, a novel spatiotemporal transformation deep learning method called Trident Segmentation CNN (TS-CNN) is proposed to segment PWML in MR images. It can convert spatial information into temporal information, which reduces the consumption of computing resources. Furthermore, a new improved training loss called Self-balancing Focal Loss (SBFL) is proposed to balance the loss during the training process. The whole model is evaluated on a dataset of 704 MR images. Overall the method achieves median DSC, sensitivity, specificity, and Hausdorff distance of 0.6355, 0.7126, 0.9998, and 24.5836 mm which outperforms the state-of-the-art algorithm. (The code is now available on https://github.com/YalongLiu/Trident-Segmentation-CNN)

*Index Terms—* Punctate white matter lesions, semantic segmentation, preterm neonates, CNN

## 1. INTRODUCTION

Punctate white matter lesions (PWML) are common in preterm neonates. It is mainly characterized by punctate, linear, or cluster hyper-signal on T1-weighted images (T1wI) and hypo-signal on T2-weighted images (T2wI) [1]. Many PWML segmentation methods are based on these characteristics. Also, the segmentation of PWML has many difficulties. First, the lesion area is much smaller than the whole brain area, so the data imbalance problem between positive and negative pixels makes it a huge challenge for deep learning models [2]. Second, the area of PWML is extremely tiny. A little segmentation deviation produced by algorithms may result in a complete segmentation failure [2]. Many works have been proposed to overcome these difficulties. Cheng et al. [3] used a stochastic process to model the neonatal brain MR images, which avoided the assumption that the features of the lesion obeyed the Gaussian distribution. But it needs to segment the white matter region manually at first. Recently, this method has been proved that it is not robust enough for MR images with low resolution and some noise [4]. Then, Mukherjee et al. [4] proposed a new method that considers the correlation of pixels in 3D space. However, they evaluated their method on a small dataset, which cannot prove the robustness of the model. To this end, a robust deep learning model on a larger data set has been proposed in our early work [2]. But its segmentation process is based on 2D slices and ignores the pixel relationships between adjacent slices. In recent years, deep learning methods especially convolutional neural network (CNN) is widely used to process medical images, like breast tumors classification [5], and brain tumor segmentation [2].

In this work, a novel and efficient PWML segmentation deep learning method called Trident Segmentation CNN (TS-CNN) is proposed. It has a unique spatiotemporal transformation structure that can utilize the information between adjacent slices. It achieves higher performance than existing methods and consumes less computing resources than 3D CNNs in theory. To better training the network, an improved loss function called Self-balancing Focal Loss (SBFL), which based on the classical Focal Loss (FL) [6] is proposed. The SBFL divides the whole loss into foreground loss and background loss part. It can automatically balance these losses and ultimately boosts the performance of the model. Finally, when compared with other existing methods, our model achieves better results in the segmentation of PWML.

[#] These authors contributed equally to this work.

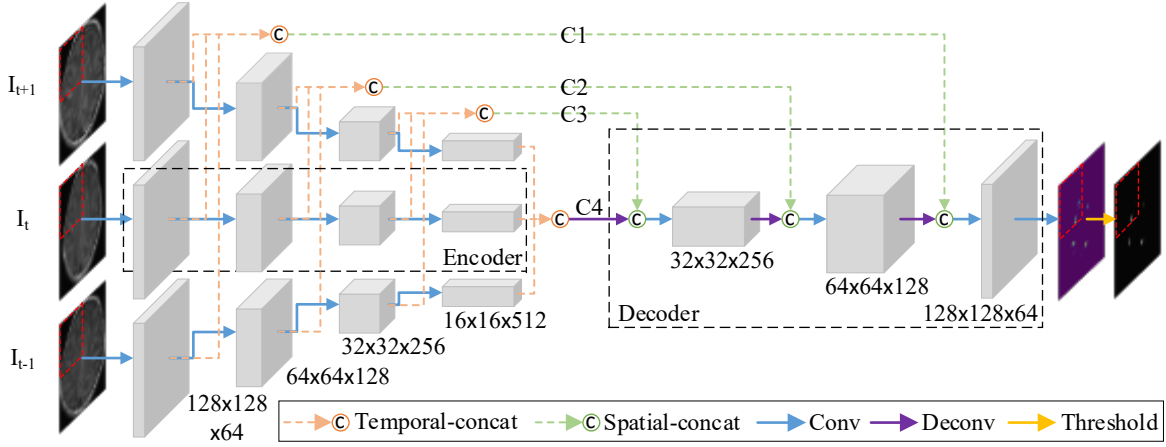

**Fig. 1**. The structure of Trident Segmentation CNN.

## 2. METHODS

### 2.1. Trident Segmentation CNN

There some correlations between the same coordinate pixels on adjacent slices as the PWML is a kind of brain tumor [4]. So, we will detect PWML on 3D voxels instead of 2D pixels for better reliability. However, if we use 3D CNN directly to deal with the entire brain image or take more than 3 slices as the input, many computing resources will be required, and it will further aggravate the class imbalance situation as the lesion area of PWML is very small when compared with the background area. Therefore, we should only consider its adjacent slices when segmenting PWML from a slice.

As is shown in Figure 1, the Trident Segmentation CNN (TS-CNN) has a classical encoder-decoder structure as residual U-Net [7]. However, there are three inputs in TS-CNN. $I_t$ is the slice which needs to be segmented while $I_{t-1}$ and $I_{t+1}$ are its adjacent slices. Firstly, to alleviate the class imbalance a little bit, we adapt the patch-based method to split the input images into small pieces half the size. Then, the TS-CNN uses a same encoder to extract the features of $I_{t-1}$, $I_t$ and $I_{t+1}$ at different times which cleverly converts a space problem to a time problem. In this way, we can improve the utilization of network parameters and reduce the consumption of computing resources at the same time. Then, features of the same layer will be concatenated on the time channel, and the time channel will be merged into the space channel by a reshape operation in the Temporal-concat module. Four different scales of feature map C1, C2, C3, and C4 will be obtained as the output of the encoder. The decoder module is a standard decoder. It extends the feature maps from the encoder with de-convolution, concatenates feature maps on the spatial channel by the Spatial-concat module, and finally produces the prediction map by a threshold segmentation. Moreover, we use deep residual units instead of plain units as basic blocks to build the model as residual U-Net [7] for better performance.

### 2.2. Self-balancing Focal Loss

In the case of unbalanced positive and negative pixel samples, the use of binary cross-entropy loss may result in the failure of deep learning model training. The Focal Loss [6] overcomes this by giving different weights to the loss of different categories. So models can be trained normally by the focal loss when the number of samples is uneven. The definition of the focal loss is described below:

$$FL = -\alpha \times y_{true} \times (1 - y_{pred})^\gamma \times \log(y_{pred} + \varepsilon) \\ - (1 - y_{true}) \times (1 - \alpha) \times y_{pred}^\gamma \times \log(1 - y_{pred} + \varepsilon), \quad (1)$$

Where $\alpha$ controls the loss ratio between positive samples and negative samples. $y_{true}$ is the ground truth, and $y_{pred}$ is the output probability map of the model. $\gamma$ is used to manage the loss ratio between hard samples and easy samples. $\varepsilon$ is a small constant used to prevent the extreme case of excessive loss value in the initial stage of training.

The essence of focal loss is to manually set the value of $\alpha$ to balance the loss between positive and negative samples. However, the ratio of the positive and negative pixels is always changing during training epochs, so there is no flexibility by manually setting the value of $\alpha$. Therefore, we propose an adaptive loss function called Self-balancing Focal Loss (SBFL). SBFL can automatically adjust the value of $\alpha$ according to the loss of positive and negative samples in the course of training. It can automatically balance the loss of positive and negative samples in most cases, and finally improve the performance of the model.

$$SBFL_0 = -(1 - y_{pred}) \times y_{pred}^\gamma \times \log(1 - y_{pred} + \varepsilon), \\ SBFL_1 = -y_{pred} \times (1 - y_{pred})^\gamma \times \log(y_{pred} + \varepsilon), \\ \beta = \frac{0.4 \times sum(SBFL_0)}{sum(SBFL_0) + sum(SBFL_1)} + 0.5, \\ SBFL = \beta \times SBFL_1 + (1 - \beta) \times SBFL_0, \quad (2)$$

Where $SBFL_0$ and $SBFL_1$ are the focal loss of background and foreground pixels without $\alpha$. $sum(SBFL_0)$ and $sum(SBFL_1)$ are the sums of $SBFL_0$ and $SBFL_1$. So $sum(SBFL_0)+sum(SBFL_1)$ represents the total loss of the model. Intuitively, we can use the ratio of $sum(SBFL_0)$ and $sum(SBFL_0)+sum(SBFL_1)$ to balance the loss of $SBFL_0$ and $SBFL_1$. However, this may result in a sharp oscillation of the model parameters. In order to segment PWML, we will focus on the segmentation of the lesion areas when balancing the loss of positive and negative samples. That is, $\beta$ should always be greater than 0.5. It also should ensure that the model does not only focus on the segmentation of positive areas, so we limit the maximum value of $\beta$ to 0.9 by a coefficient of 0.4. Finally, the $SBFL$ is composed of $SBFL_1$ weighted by $\beta$ and $SBFL_0$ weighted by $1-\beta$.

### 2.3 Implementation Details

We implement the proposed method in Keras of TensorFlow backend. All the models are trained and tested on an NVIDIA 1080Ti GPU. The optimizer is Adam with no weight decay. The learning rate is 2e-4, and the batch size is 8. The training epochs and the steps per epoch are 50.

### 3. EXPERIMENTS AND RESULTS

#### 3.1. Datasets and Preprocessing

The T1w MR images from 70 neonates enrolled in this paper are provided by the local hospital with a 3T MRI scanner. There are 704 slices with PWML, and the size of each slice is 256 × 256 pixels. To avoid unnecessary calculation and remove significant interference information, we crop the excess part of the image and retain only the square area of the skull and its inner space. These images are upsampled to 256 × 256 pixels as the input image $I$.

#### 3.2. Results

The performances of different models are shown in Table 1. Row A is the traditional PWML segmentation method [4] proposed by Mukherjee et al. in 2019. We use the residual U-Net [7] to segment PWML in row B. Row C is our previous method called RS-RCNN [2]. Row D is the proposed model. It gets better performance than RS-RCNN [2], which proves that the theory and the structure of TS-CNN are very reliable. Then we apply another data augmentation during the model training process and get a little improvement in all indexes as shown in row E. To enhance the feature extraction ability of our model, we double its channel counts (CC) to 64, and the results are list in row F. Row B to row F are all trained by focal loss. Finally, we apply the SBFL to the model in row F and get a huge performance improvement as shown in row G that approves the effectiveness of SBFL.

**Table 1**. The performance of PWML segmentation models. All these results are the medians of 5-folds cross-validation.

| | Model | DSC | Sensitive | Specificity | Hausdorff Distance |
|---|---|---|---|---|---|
| A | Mukherjee et al. [4] | 0.4288 | 0.4533 | 0.9961 | 59.6432 |
| B | Residual U-Net [7] | 0.5838 | 0.6013 | **0.9998** | 44.7323 |
| C | RS-RCNN [2] | 0.5986 | 0.6535 | **0.9998** | 36.6065 |
| D | TS-CNN | 0.6024 | 0.6838 | **0.9998** | 30.9768 |
| E | TS-CNN (Data-aug) | 0.6088 | 0.6838 | **0.9998** | 30.8133 |
| F | TS-CNN (CC=64) | 0.6186 | 0.6952 | **0.9998** | 28.3413 |
| G | TS-CNN (SBFL) | **0.6355** | **0.7126** | **0.9998** | **24.5836** |

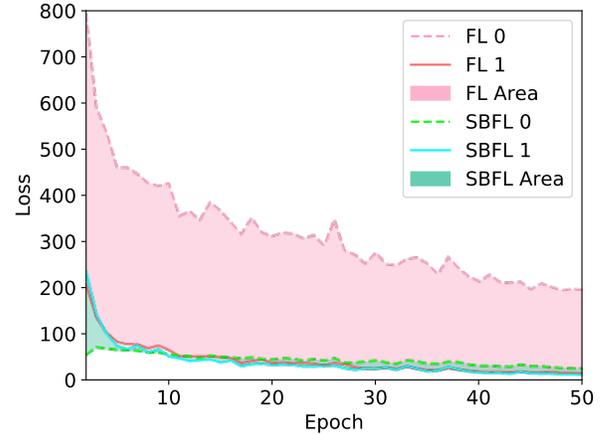

**Fig. 2**. The effect of focal loss and SBFL on the model training loss.

For better understanding the mechanism of SBFL. We visualize the loss value and the difference between positive sample loss and negative sample loss of Focal Loss (FL) [6] and SBFL in Figure 2. The pink dashed line and the red solid line represent the background loss and foreground loss respectively in the focal loss while the green dashed line and the blue solid line are the losses in SBFL. The pink area is the difference between $FL_0$ and $FL_1$ in the focal loss, while the green area is the difference between $SBFL_0$ and $SBFL_1$ in SBFL. We set the $\alpha$ to 0.9 in focal loss, but there is still a huge difference between $FL_0$ and $FL_1$. Intuitively, we can further increase $\alpha$ until $FL_0$ is close to $FL_1$, but the ratio of positive and negative pixels is continually changing while $\alpha$ is fixed during training. The SBFL can automatically compute the value of $\beta$ by the positive and negative sample losses. It can automatically focus on the positive sample loss during the initial stage of the training and automatically balance the positive and negative sample losses during the subsequent training phases. The value of $FL_1$ is very close to the value of $SBFL_1$ which indicates that SBFL algorithm can automatically decay the background prediction error and has no negative influence on the foreground prediction accuracy at the same time.

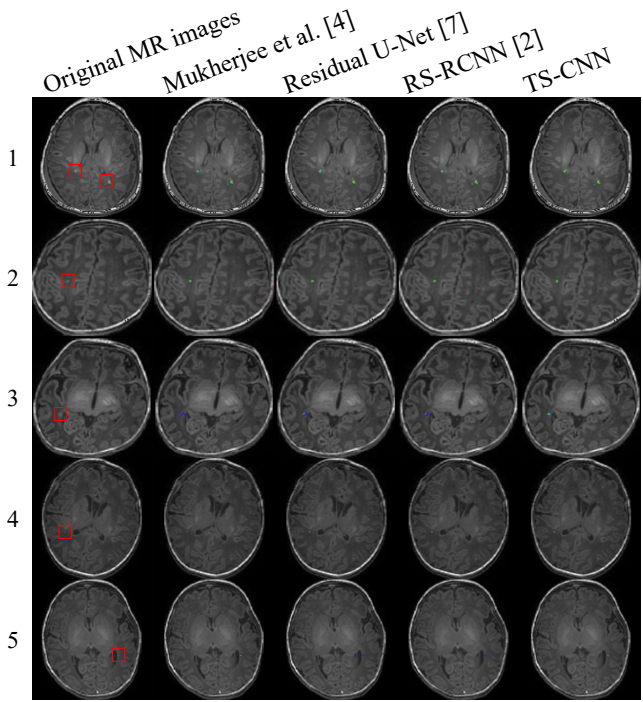

**Fig. 3**. The visualization of segmentation results of different models. PWML is located in red squares in the original images. The red pixels are false positive (FP) while blue is false negative (FN), and green is true positive (TP) in the predictions of four different models.

We show the PWML segmentation results of different models in Figure 3. In row 1, the area of the lesion is scattered but all lesions are detected by four methods. The FN predictions in TS-CNN are much less, and there is a false positive lesion detected by Mukherjee et al. [4]. In row 2, there are many interference areas that look like PWML. TS-CNN can well distinguish them by the information from adjacent slices, while there are many wrong detections made by other methods. In row 3, the tumor has no distinct boundary. Only TS-CNN can detect nearly the whole tumor while other methods only figure out a small part of the lesion. The PWML is not visually apparent in row 4, but both Residual U-Net [7] and TS-CNN get good results. In row 5, the features of PWML are very inconspicuous and there is a brain hemorrhage in the lower part of the image. Both Mukherjee et al. [4] and TS-CNN detected the lesion while the brain hemorrhage is identified as a PWML by the former method. So, we believe that TS-CNN has robust PWML detection and anti-interference ability.

## 4. CONCLUSION

In this paper, we propose the TS-CNN and the Self-balancing Focal Loss for better-segmenting PWML. The TS-CNN is an efficient deep learning model that can utilize the information from adjacent slices in MR images by transforming spatial information into temporal information and finally get better results than other existing methods. The Self-balancing Focal Loss is an improved version of focal loss. It can automatically balance the loss between positive samples and negative samples in the model training process and significantly boost the model performance. We believe that our TS-CNN and Self-balancing Focal Loss have practical implications for other studies.

## 5. ACKNOWLEDGEMENTS

This research was supported in part by the National Natural Science Foundation of China under Grant 61671339, 61432014 and 61772402, and in part by National High-Level Talents Special Support Program of China under Grant CS31117200001. Yalong Liu, Miaomiao Wang, and Xianjun Li contributed equally to this work. Please address correspondence to Xianjun Li (xianj.li@mail.xjtu.edu.cn).